\documentclass[epsf, preprint2]{aastex}
\shorttitle{Nature of eclipsing pulsars}
\shortauthors{Khechinashvili, Melikidze, and Gil}
\begin{document}
\title{Nature of eclipsing pulsars}
\author{David G. Khechinashvili\altaffilmark{1,2}, George I. Melikidze\altaffilmark{1,2}, and Janusz A. Gil\altaffilmark{1}}
\email{dato@astro.ca.sp.zgora.pl, gogi@astro.ca.sp.zgora.pl, jag@astro.ca.sp.zgora.pl}
\altaffiltext{1}{J. Kepler Astronomical Center, Pedagogical University, Lubuska 2, 65-265 Zielona G\'ora, Poland}
\altaffiltext{2}{Abastumani Astrophysical Observatory, Al. Kazbegi ave. 2a, 380060 Tbilisi, Georgia}
 
\begin{abstract}
We present a model for pulsar radio eclipses in some binary systems, and test this model for PSRs B$1957+20$ and J$2051-0827$. We suggest that in these binaries the companion stars are degenerate dwarfs with strong surface magnetic fields. The magnetospheres of these stars are permanently infused by the relativistic particles of the pulsar wind. We argue that the radio waves emitted by the pulsar split into the eigenmodes of the electron-positron plasma as they enter the companion's magnetosphere and are then strongly damped due to cyclotron resonance with the ambient plasma particles. Our model explains in a natural way the anomalous duration and behavior of radio eclipses observed in such systems. In particular, it provides stable, continuous, and frequency-dependent eclipses, in agreement with the observations. We predict a significant variation of linear polarization both at eclipse ingress and egress. In this paper we also suggest several possible mechanisms of generation of the optical and $X$-ray emission observed from these binary systems.
\end{abstract}
\keywords{binaries: eclipsing --- pulsars: individual (PSR B$1957+20$, J$2051-0827$)}

\section{Introduction \label{Introduction}}
Eclipsing millisecond pulsars were expected to be a missing link in the evolutionary connection between millisecond pulsars and low-mass $X$-ray binaries \citep{alpar82}. To the best of our knowledge, the following eclipsing millisecond pulsars have been observed for today: PSR B$1957+20$ \citep{fst88}, PSR B$1744-24$A \citep{lyne90}, PSR B1259-63 \citep{johnston92}, PSR J$2051-0827$ \citep{stappers96}, 47~Tuc~I,~J \citep{man91}, O, and~R \citep{c99}. Up to now, PSRs B$1957+20$ and J$2051-0827$ are the best studied of them. These two pulsars exhibit stable, anomalously long and frequency-dependent radio eclipses, that is, eclipses are long at low observed frequencies, and become shorter at higher frequencies. 

In this paper we propose a physical mechanism for the eclipsing binary pulsars and test it for PSRs B$1957+20$ and J$2051-0827$. Our model assumes that the companion stars are magnetic white dwarfs. Relativistic particles from the pulsar wind are trapped by the companion's magnetic field. This leads to formation of the extended magnetosphere of the companion star. We demonstrate that the pulsar radio emission undergoes strong cyclotron damping while passing through the companion's magnetospheric plasma. The model accounts for the most peculiarities of the observed eclipse pattern in both aforementioned eclipsing systems. To begin with, let us shortly review the main observational features of these interesting objects.

\subsection{PSR B$1957+20$ \label{PSR B$1957+20$}}
One of the two fastest pulsars known (with the period $P\approx 1.6$~ms and the spin-down rate $\dot{P}\approx 1.7\times 10^{-20}$), the ``black widow'' pulsar B$1957+20$ shows regular and entirely periodic eclipses at meter wavelengths, which occupy about 10\% of its $T_{\mathrm{orb}}\approx 9.2$~hr orbital period. The eclipses are quite stable in length at any given observing frequency, although their length depends strongly on the frequency: at 318~MHz it averages to about 55 minutes, but decreases to about 33 minutes at 1.4~GHz. This frequency dependence appears to fit a $\Delta t_{e}\propto \nu^{-0.4\pm 0.1}$ power law, where $\Delta t_{e}$ denotes eclipse duration at a frequency $\nu$. The orbit of the binary system is rather circular, with the separation of the companions $a\approx 1.7\times 10^{11}~\mathrm{cm}\approx 2.5\,R_{\odot }$ \citep{bt95}, where $R_{\odot }=6.96\times 10^{10}$~cm is a solar radius.

At lowest observing frequencies eclipse ingress is quite rapid, whereas eclipse egress is slower and somewhat turbulent. At 1.4 GHz, however, this asymmetry is not observed. The excess electron density on either side of the eclipse is found to vary by a factor of two from eclipse to eclipse. The delay between the left and right circularly polarized pulse arrival times were also detected \citep{tfst89,fbb90,rt91}.

HST optical images of the PSR B$1957+20$ made at different orbital phases display a dramatically variable system which is brightest when the side of the companion heated by the pulsar wind faces the Earth, and darkens by several magnitudes when the opposite cool side comes into view \citep{frucht95}. Companion's radius, $R_{\mathrm{opt}}\approx 0.12\,R_{\odot}$, was derived from the optical images using the estimated distance to this object $d=0.8$~kpc. Therefore $R_{\mathrm{opt}}$ is less than one-half the Roche lobe radius $R_{L}\approx 0.29\,R_{\odot}$, and the enigmatic companion of PSR B$1957+20$ does not even fill its Roche lobe \citep{de88}.

Soft $X$-rays from this system were detected using the ROSAT observatory \citep{fg92,kpeh92}. An $X$-ray luminosity of about $10^{31}~\mathrm{erg~s^{-1}}$ (which is $\sim 10^{-4}$ of the pulsar spin-down energy) was found, although no variability has been detected. Below we discuss possible sources of this high-energy and optical radiation.

\subsection{PSR J$2051-0827$ \label{PSR J$2051-0827$}}
The discovery of this eclipsing millisecond binary system was reported by \citet{stappers96} who carried out timing observations at several frequencies between 408 MHz and 2.0 GHz. The orbital period $T_{\mathrm{orb}}\approx 2.38$~hr of this system makes it the third shortest pulsar orbital period known, behind 47~Tuc~R \citep{c99} and PSR B1744-24A \citep{lyne90}. Thus, the system is extremely compact, with a separation of the binary components $a\approx 1.0\,R_{\odot}$. High-precision timing observations of PSR J$2051-0827$ (the dynamical parameters of this pulsar are: $P=4.5$~ms and $\dot{P}=1.3\times 10^{-20}$) indicate that the orbital period is decreasing at a rate of $\dot{T}_{\mathrm{orb}}=\left( -11\pm 1\right)\times 10^{-12}$ \citep{sbmst98}. Such an orbital period derivative implies a decay time for the orbit of only $25$~Myr, which is much shorter than the expected timescale for the ablation of the companion. This, in combination with a very slow transverse velocity of the pulsar itself, questions the formation of the system in the standard manner \citep{sbmst98}.

The duration of the pulsar eclipse at 436 MHz is $\sim 10\%$ of the orbital period, which implies a radius of the eclipse region surrounding the companion $R_{e}\approx 2\times 10^{10}~\mathrm{cm}\approx 0.3\,R_{\odot}$. The variation in eclipse duration between 436 and 660 MHz seems to be frequency dependent with $\Delta t_{e}\propto\nu^{-0.15}$. On the other hand, at 1.4 GHz the pulsar emission was detected throughout the low-frequency eclipse region in approximately half of the observing sessions \citep{stappers96}. These results, along with the observations at 2.0~GHz, suggests that no eclipse occurs at high radio frequencies.

\citet{stapgb96} also detected an optical counterpart of the companion of PSR J$2051-0827$. The amplitude of its light curve is at least 1.2~mag. Thus, PSR J$2051-0827$ is the second pulsar binary for which direct heating of the companion was observed. The radius of the companion's optical counterpart (derived in the blackbody approximation) falls in the range $R_{\mathrm{opt}}\sim 0.067 -  0.18\,R_{\odot}$ \citep{stapgb96}. Thus, the companion star is likely to fill its Roche lobe ($R_{L}\approx 0.13\,R_{\odot}$). The best-fit models, based on the new photometry of the secondary star \citep{sklk99}, require that greater than 30\% of the incident energy is absorbed by the companion and reradiated as optical emission. The unilluminated side of the companion (i.e., observed at an orbital phase corresponding to the pulsar radio eclipse) was found to be cool \citep{sklk99}, with a best-fit temperature of 3000~K, similar to that obtained for the companion to PSR 1957+20 \citep{fbb95}.

\section{Nature of the companion stars \label{Nature of companions}}
Let us begin the discussion with a few comments on the nature of the companion stars. The minimum masses of the companions in both cases were calculated from the mass function $f(m_{p},m_{c})$ using observed orbital parameters and assuming the pulsar mass $m_{p}\approx 1.4\,M_{\odot }$ and the orbital inclination $i\geq 60^\circ$, the latter necessary for the existence of eclipses. For the companions of PSRs J$2051-0827$ and B$1957+20$ this yields $m_{c}\approx 0.027\,M_{\odot}$ \citep{stappers96} and $m_{c}\approx 0.022\,M_{\odot}$ \citep{fst88}, respectively (i.e., very low companion masses). The actual masses of companions are within a factor of 1.2 of the above minimum values (because of $i\geq 60^\circ$). \citet{frucht95} argues that the companion of PSR 1957+20 is obviously well below the hydrogen burning limit, and one would expect this star to be a degenerate dwarf. We suggest that the same is valid for the companion of PSR J$2051-0827$. The radius of a low-density white dwarf (assuming nonrelativistic electron Fermi gas in its interior) is given by the mass-radius relation \citep{shapteuk83} 
\begin{equation} 
\frac{R_{c}}{R_{\odot }}\approx 4.05\times 10^{-2}\left( \frac{m_{c}}{M_{\odot }}\right) ^{-1/3}\mu _{e}^{-5/3},  \label{wd-radius}
\end{equation}
where $\mu_{e}=A/Z$ is a mean molecular weight per electron\footnote{For brevity, hereafter subscripts ``1'' and ``2'' correspond to the physical quantities regarding the companions of PSRs B$1957+20$ and PSR J$2051-0827$, respectively.}. Substituting the masses of each companion star in equation~(\ref{wd-radius}) we can calculate their radii for the cases of pure hydrogen ($\mu_{e}=1$) and pure helium ($\mu_{e}=2$) white dwarfs, which yield $R_{c1}^\mathrm{H}=0.145\,R_{\odot}$, $R_{c1}^\mathrm{He}=0.046\,R_{\odot},$ $R_{c2}^\mathrm{H}=0.135\,R_{\odot}$, and $R_{c2}^\mathrm{He}=0.043\,R_{\odot}$, respectively. For a mixture of 25\% He and 75\% H by mass $(X=0.25)$, we obtain that $R_{c1}\approx R_{c2}\approx 0.1\,R_{\odot }$. We see that these values of companion radii are comparable to those estimated from the optical observations (see Section~\ref{Introduction}).

Because magnetic flux is conserved in gravitational collapse, the stellar magnetic fields can be amplified to more than a million Gauss when a star contracts to the white dwarf stage. Observations show that the magnetic fields at white dwarf surfaces vary from a few tens to millions of Gauss \citep{lang92}. We assume that both white dwarfs discussed above possess significant magnetic fields at their surfaces, and estimate these magnetic fields below.

\section{Eclipse mechanism \label{Eclipse mechanisms}}
\subsection{Review}
\citet{thompson94} and \citet{thompson95} analyzed a variety of physical mechanisms which can cause pulsar eclipses. They considered eclipses by a wind from the stellar companion, by a stellar magnetosphere, or by material entrained in the pulsar wind. The refractive eclipse proposed by \citet{pebk88} was ruled out by \citet{thompson94} due to inability of this mechanism to explain a sensitive dependence of the eclipse duration on frequency, as well as measured time delays. The other mechanisms discussed by \citet{thompson94} and \citet{thompson95} are as follows: a free-free absorption, scattering by plasma turbulence, induced Compton scattering, stimulated Raman scattering parametric instability and synchro-cyclotron absorption. \citet{thompson94} and \citet{thompson95} are inclined to the opinion that the eclipse is due to cyclotron absorption by plasma of temperature $T_{e}\sim (1 - 4)\times 10^{8}$ K and density $n_{e}\simeq 4\times 10^{4}$ cm$^{-3}$ in a field of strength $B_{e}\sim 15 - 20$ G. Another possibility is a synchrotron absorption by relativistic plasma electrons \citep{eichler91}.

Detailed discussion of eclipse mechanisms was also presented by \citet{eichged95}. They focus on three wave processes, in particular Raman scattering (the decay $photon\rightarrow photon+plasmon$ stimulated by an ambient plasmon field) and induced plasmon emission (i.e. the same process stimulated by photons already in the final state). These authors claim that both mechanisms are able to produce pulsar eclipses via large angle scattering on the initially beamed pulsar radiation, and neither of them requires high plasma densities. \citet{bt95} considered different scenarios of the mass outflow in binaries, such as mass outflow intrinsic to the companion star and pulsar irradiation-driven outflows or Roche lobe overflow, and discarded the latter.

A mechanism based on three-wave interactions involving low-frequency acoustic turbulence was proposed to be responsible for pulsar eclipses by \citet{lm95,lc97}. The authors claim that the eclipsing material consists of plasmas from the companion wind, heated by the pulsar wind. This may result in a hot corona-like plasma which is possibly non-isothermal. They considered both the case of electron acoustic waves when the ion acoustic temperature $T_{i}$ is higher than the electron temperature $T_{e}$ and that of ion acoustic waves when $T_{e}>T_{i}$. It was found by \citet{lc97} that induced scattering off electron acoustic waves can be important and can cause pulsar eclipse. At the same time, it was demonstrated that induced Brillouin scattering involving low-frequency ion acoustic waves cannot be the main reason of eclipse, because of the relatively small growth rates of these waves.

\subsection{Geometry of the model \label{Geometry}}
As we have seen in Section~\ref{Introduction}, the ratio of the eclipse duration (at low frequenices) to the orbital period is about the same in both PSR B$1957+20$ and PSR J$2051-0827$, i.e., $\Delta t_{e}/T_{\mathrm{orb}}\approx 0.1$. As the eclipse nears, the pulse decreases in amplitude but does not change in shape, showing that the eclipse mechanism removes photons from the line of sight but does not significantly scatter the pulse. Thus, only absorption or large-angle scattering may be responsible for it \citep{stappers96}. On the other hand, the eclipse length indicates that the eclipsing medium extends to the distances much larger than the dimensions of a companion of any conceivable composition \citep{frucht95}. Indeed, for PSR B$1957+20$ the radius of eclipsing region $R_{e}(318\,\mathrm{{MHz}})\approx 0.76\,R_{\odot}$ (so that the ``eclipsing spot'' lies at least $1.5\,R_{\odot }$ across), $R_{e}(1.4\,\mathrm{GHz})\approx 0.46\,R_{\odot },$ whereas for PSR J$2051-0827$ $R_{e}(436\,\mathrm{{MHz}})\approx 0.31\,R_{\odot }$. We see that in both cases (at all radio frequencies) the eclipse region lies even well beyond the corresponding companions' Roche lobes ($R_{L1}\approx 0.29\,R_{\odot }$ and $R_{L2}\approx 0.13\,R_{\odot}$, respectively). 

\citet{thompson95} points out that the large size of the eclipses (compared to the companion's Roche lobe) could be explained either by diffusion of plasma into the pulsar wind, or by a large companion magnetosphere. We adopt this statement and claim that companions of both pulsars are degenerate dwarfs with significant magnetic fields at their surfaces. At the same time, we suggest that the magnetospheres of these magnetic white dwarfs are infused with relativistic particles supplied permanently by the pulsar relativistic wind. Figure~\ref{eclipse-geometry} represents schematically geometry of such a binary system.

\vskip 12pt~\scalebox{0.5}{\includegraphics{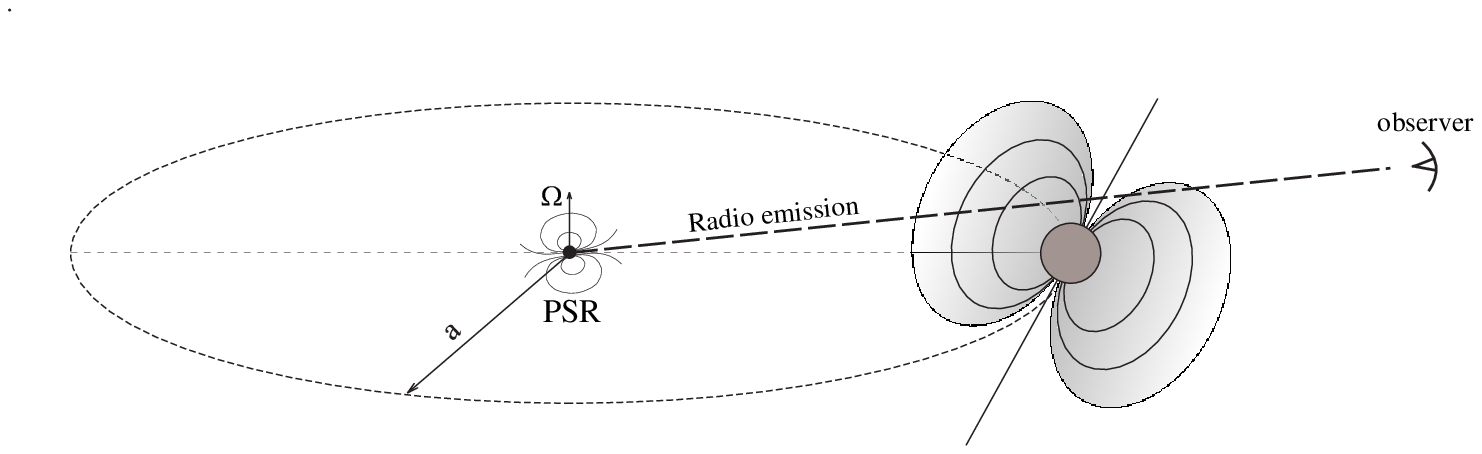}}~\figcaption[f1.eps]{Schematic representation of a binary system (viewed at some angle) at an orbital phase corresponding to the radio eclipse. $a$ is the binary separation. An extended magnetosphere of the white dwarf is continuously supplied by relativistic plasma particles of the pulsar wind.\label{eclipse-geometry}}~\vskip 2pt

Let us estimate the density of a wind plasma at the distances from the pulsar surface corresponding to the binary separation $a$. The number density of pair plasma inside the pulsar light cylinder (with the radius $R_{_\mathrm{LC}}=Pc/2\pi$) is
\begin{equation}
n_{_p}=2.2\times 10^{18}\frac{\kappa }{\sin \alpha }\left( \frac{\dot{P}}{P}\right)^{0.5}\left( \frac{R_{0}}{R}\right)^{3}~\mathrm{[cm^{-3}]}, \label{n_lc}
\end{equation}
where $\kappa$ is a Sturrock multiplication factor \citep{s71} and $\alpha$ is an inclination angle between the pulsar magnetic and rotation axes. From equation~(\ref{n_lc}) we can calculate the number density of particles at distance of one light cylinder radius, $n_{_{pL}}$. Assuming further that the density falls according to the inverse square law beyond the pulsar light cylinder $n_{_p}(R)={n_{_{pL}}}(R_{_\mathrm{LC}}/R)^{2}$ we get 
\begin{equation}
n_{_p}(R)=3\times 10^{-3}\frac{\kappa}{\sin \alpha}\left( \frac{\dot{P}_{-15}}{P^3}\right)^{0.5}
\left( \frac{R_{\odot}}{R}\right)^2~\mathrm{[cm^{-3}]}.\label{n_ecl}
\end{equation}
Substituting in equation~(\ref{n_ecl}) the dynamic parameters of each of the pulsars discussed in this paper, taking $\alpha \approx 90^\circ$ for PSR B$1957+20$ (note that this pulsar has an interpulse indicating that it is an almost orthogonal rotator), and assuming that the inclination angle is rather large for PSR J$2051-0827$ too, we find that at the distances corresponding to binary separations in each of the systems ($a_{1}\approx 2.5\,R_{\odot}$ and $a_{2}\approx 1.0\,R_{\odot}$, respectively), the plasma densities in pulsar winds are $n_{_p}(a_{1})\approx 3.1\times 10^{-2}\kappa$~cm$^{-3}$ and $n_{_p}(a_{2})\approx \,3.6\times 10^{-2}\kappa~\mathrm{cm}^{-3}$, respectively. We see that the densities are about the same order of magnitude in these two cases. Note that the actual value of plasma density in the companions' magnetospheres may be much higher than inferred from equation~(\ref{n_ecl}). The point is that the slowest particles of pulsar wind trapped by the companion's magnetic field should keep bouncing between its magnetic poles without relative density loss on an extended timescale. Such an accumulation of plasma particles may apparently increase their number density.

\subsection{The model}
We share the opinion expressed by \citet{thompson94} that the most promising model of eclipsing pulsar binaries is cyclotron absorption. However, we believe that the proper consideration of waves damping in the relativistic magnetized plasma should be performed basing on kinetic treatment of the cyclotron damping. Basing on this statement, as well as on the physical picture of an eclipsing binary system presented in Section~\ref{Geometry} (see also Fig.~\ref{eclipse-geometry}), we presented a model for the radio eclipse of PSR B$1957+20$ \citep{km97}. Here we develop and further extend this model by applying it to PSR J$2051-0827$.

Let us check the possibility of waves damping in the plasma of companion's magnetosphere. Radio waves emitted by the pulsar (with the vacuum spectrum $\omega=kc$) should split into the eigenmodes of the relativistic pair plasma, as they enter the companion's magnetosphere. It was shown \citep[e.g.,][]{lmmp86,ab86} that in the magnetized relativistic pair plasma there exist three wavemodes with the frequencies much less than $\omega_{_B}/\gamma_{_p}$, where $\omega_{_B}=eB/mc$ is the gyrofrequency and $\gamma_{_p}$ is a mean Lorentz factor of the plasma particles. In the general case of oblique propagation with respect to the local magnetic field, these eigenmodes are: i) the purely electromagnetic extraordinary (X) mode; ii) the subluminous Alfv\'{e}n (A) mode; and iii) the superluminous ordinary (O) mode. The last two modes are of mixed electrostatic-electromagnetic nature. Electric field vectors ${\mathbf E}^{\rm O}$ and ${\mathbf E}^{\rm A}$ of O and A-modes lie in the $\left( {\mathbf k}\,{\mathbf B} \right)$ plane, while the electric field of X-mode ${\mathbf E}^{\rm X}$ is directed perpendicularly to this plane.

All the three wavemodes may be strongly damped in the companion's magnetosphere at the cyclotron resonance
\begin{equation}
\omega-k_{\parallel }v-\frac{\omega_{_B}}{\gamma_{_\mathrm{res}}}=0  \label{normal cyclotron}
\end{equation}
on the particles of ambient plasma with the mean Lorentz factor $\gamma_{_\mathrm{res}}=\gamma_{_p}$. Note that in equation~(\ref{normal cyclotron}) $k_{\parallel}$ denotes the component of the wavevector along the local magnetic field, and we neglected the curvature drift of the bulk plasma particles across the plane of the magnetic field line curvature.

Let us calculate the characteristic frequency of damping of the plasma eigenmodes at the cyclotron resonance. We assume that all the modes are vacuum-like in the domain of their spectrum where the cyclotron damping occurs, i.e. their spectrum can be represented as
\begin{equation}
\omega=kc (1-\delta),\quad \mathrm{where}\quad \delta\ll 1. 
\label{damped modes}
\end{equation}
Using the dispersion relation in this form we can rewrite the resonance condition~(\ref{normal cyclotron}) as follows 
\begin{equation}
kc(1-\delta)-kc\cos \theta \left( 1-{\frac{1}{{2\gamma_{_p}^{2}}}}\right) ={\frac{\omega_{_B}}{\gamma_{_p}}},
\label{normal cyclotron 1}
\end{equation}
where $\theta$ is an angle between the wave-vector and the magnetic field in the resonance region, and we used $v\approx 1-1/2{\gamma_{_p}^{2}}$. From equation~(\ref{normal cyclotron 1}) we obtain two approximations of the frequency of damped waves:
\begin{equation}
\omega_{_d} \approx 2\gamma_{_p}\omega_{_B},\quad{\rm if}\quad {\theta \ll}\frac{{1}}{{\gamma_{_p}}}{,}  \label{df1}
\end{equation}
and
\begin{equation}
\omega_{_d} \approx \frac{\omega_{_B}}{{\gamma_{_p}}\left( {1-\cos\theta }\right) },\quad{\rm if}\quad {\theta \gg }\frac{{1}}{{\gamma_{_p}}}{.}  \label{df2}
\end{equation}

The decrements of cyclotron damping for the plasma eigenmodes are obtained by solving the corresponding dispersion equations \citep[see, e.g.,][]{lmmp86,k99} for the imaginary part of the complex frequency ($\omega \rightarrow \omega+i\Gamma$). Therefore the decrements of X, O and A-modes write
\begin{equation}
\Gamma_{\mathrm{X}} =-{\frac{\pi }{2}}{\frac{\omega_{_p}^{2}}{\gamma_{_T}\omega_{_d}}}  \label{Decr_X}
\end{equation}
and
\begin{equation}
\Gamma_{\mathrm{O}} \approx \Gamma_{\mathrm{A}}=-{\frac{\pi }{2}}{\frac{\omega_{_p}^{2}}{\gamma_{_T}\omega_{_d}}}g\left( \theta \right),\label{Decr_OA}
\end{equation}
respectively, where 
\begin{equation}
g\left( \theta \right) =\cos ^{2}\theta +\frac{2\sin ^{2}\theta \cos \theta}{1-\cos \theta + \left(1/2\gamma_{_p}^{2}\right)},  \label{g(theta)}
\end{equation}
and $\gamma_{_T}$ is a thermal spread of the particles energy. In equations~(\ref{Decr_X}) and (\ref{Decr_OA}) we used the mean value theorem and the normalization of the distribution function, namely that $f\left(\gamma_{_p}\right) \gamma_{_T}=\int f\left( \gamma \right) d\gamma =1$. Note that for small angles (or small Lorentz factors of the ambient plasma particles) $g\left(\theta\right)\approx1$ and the decrement of O and A-modes (eq.~[\ref{Decr_OA}]) transforms into that of X-waves (eq.~[\ref{Decr_X}]), 
\begin{equation}
\Gamma_{\mathrm{(O,A)}}\approx -{\frac{\pi }{2}}{\frac{\omega_{_p}^{2}
}{\gamma_{_T}\omega_{_d}},\quad }{\rm if}\quad {\theta \ll 1/\gamma_{_p}.}
\label{Decr_OA_small}
\end{equation}
\vskip 12pt~\scalebox{0.6}{\includegraphics{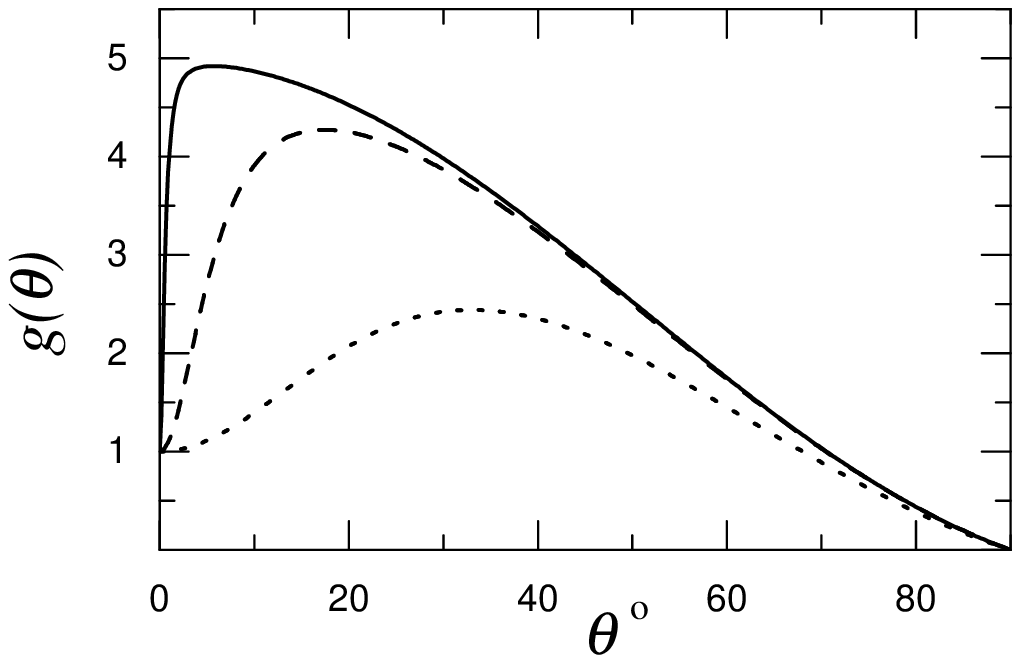}}~\figcaption[f2.eps]{Angular dependence of the decrement of O and A-modes in the limit ${\gamma_{_p}\theta \gg 1}$. The function $g(\theta)$ defined by equation~(\ref{g(theta)}) is plotted versus $\theta $ in degrees, for three different values of the mean Lorentz factor -- $\gamma_{_p}=2$, $\gamma_{_p}=10$ and $\gamma_{_p}=100$ -- represented by the dotted, dashed and solid lines, respectively.\label{g(theta)-plot}}~\vskip 2pt
In the opposite limit, i.e. when the angles (or the plasma mean Lorentz factors) are relatively large, $g(\theta)$ is plotted in  Figure~\ref{g(theta)-plot}, for different values of $\gamma_{_p}$. It is seen that if ${1/\gamma_{_p}\ll \theta <1}$ the value of the decrement of O and A-modes is a few times larger than that of the X-mode (provided that the other parameters remain the same in both cases). For example, for $\theta =30^\circ$ and $\gamma_{_p}=100$ (a dashed line in Fig.~\ref{g(theta)-plot}) we have $\Gamma_{\mathrm{(O,A)}}/\Gamma_{\mathrm{X}}\approx 4$. The two decrements become equal each other $\left( g(\theta )\approx 1\right)$ for the angle $\theta \approx 70^\circ$. On top of that, it appears that O and A-modes propagating nearly
transversely to the local magnetic field undergo no cyclotron damping, since $g\left( \theta \right) \rightarrow 0$ if $\theta \rightarrow 90^\circ$. Indeed, the A-mode vanishes at exactly transverse propagation to the local magnetic field (its group velocity tends to zero), whereas the O-mode transforms into a mode whose electric field is parallel to the external magnetic field and ${\mathbf k}\perp {\mathbf E}$. Such a wave is not subject to cyclotron damping.

It is obvious that the angles between wavevectors and the local magnetic field cannot be small, hence, $\theta \gg 1/\gamma_{_p}$. Therefore, we can calculate the frequency of damped waves from equation~(\ref{df2}), which is now convenient to represent in the following form 
\begin{equation}
\frac{\nu_{_d}}{\mathrm{[GHz]}}\approx 2.8\times 10^{-3}\frac{B_{c}}{{\gamma_{_p}}\left( {1-\cos \theta }\right) }\left( \frac{R_{c}}{r}\right)^{3}.  \label{nu_d_eclipse}
\end{equation}
Here $r$ is a distance from the companion's center, $R_{c}$ is the radius of the companion star, and $B_{c}$ is its surface magnetic field measured in Gauss. Note that we assume the companion's magnetic field to be dipolar, i.e. $B\left( r\right)=B_{c}(R_{c}/r)^{3}$.

\vskip 12pt~\scalebox{0.55}{\includegraphics{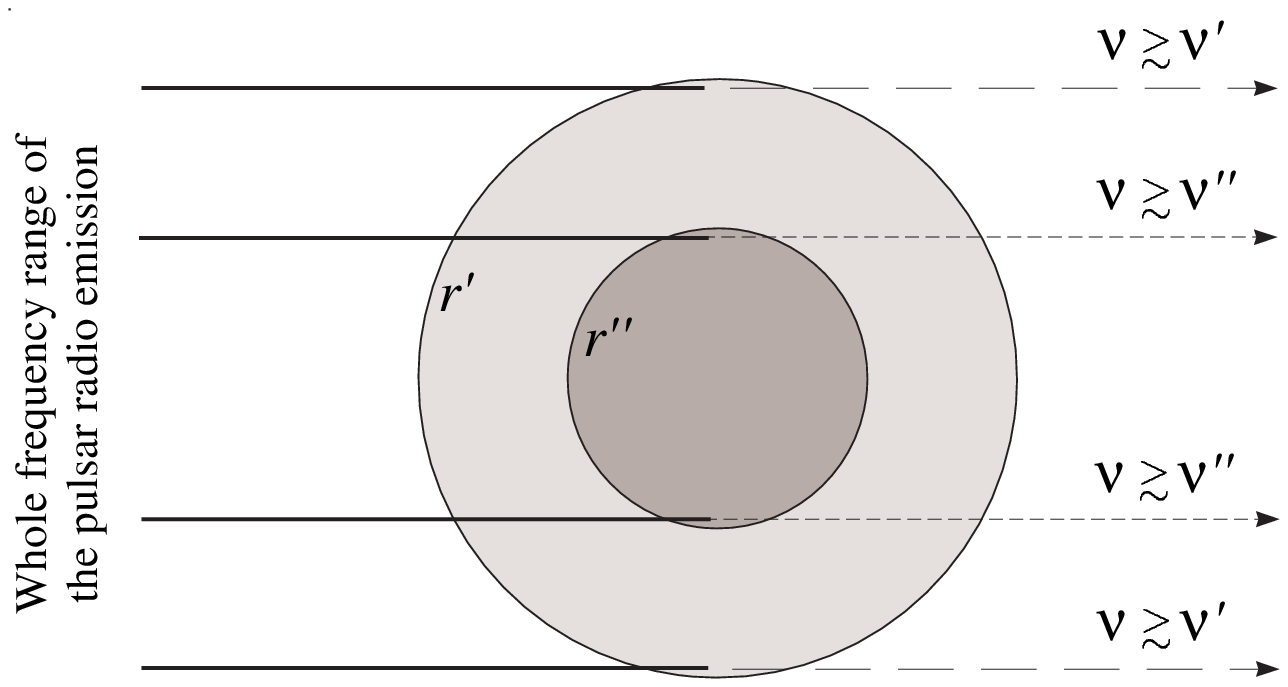}}~\figcaption[f3.eps]{``Eclipsing spots'' at different frequencies. The pulsar emission reaching the damping region represents a full range of radio frequencies. It is seen that all the frequencies lower than $\nu'$ are damped at distances from the stellar surface larger than $r'$. The waves with higher frequency $\nu''>\nu'$ propagate freely through this outer region with a low value of the magnetic field strength $B'$. However, they are damped at the distance $r''$ where magnetic field value is higher, $B''>B'$. All the radio waves with frequencies $\nu\geq\nu''$ escape this narrow inner region. \label{eclipse-spots}}~\vskip 2pt

It is seen from equation~(\ref{nu_d_eclipse}) that the frequency of damped waves is inversely proportional to the cube of the distance from the companion's surface (obviously, this dependence results from $\nu_{_d}\propto \omega_{_{Be}}\propto B_{e}$, the latter being the magnetic field value in the eclipse region). In other words, distinct frequencies are damped at distinct heights above the stellar surface (corresponding to appropriate values of the magnetic field), while higher frequencies are damped closer to the companion star. The waves with frequencies somewhat higher and lower than the one given by equation~(\ref{nu_d_eclipse}) propagate almost freely in the plasma at this distance from the star. Although, they can also reach a region at a different altitude or with different direction of local magnetic field (hence, angle $\theta $) and be damped there. Figure~\ref{eclipse-spots} represents schematically a top view of the eclipse region. It follows from this picture that the size of ``eclipsing spot'' is frequency-dependent, being larger at lower frequencies. Indeed, radio waves with a low frequency $\nu'$ (from the entire range of the pulsar radio spectrum) are damped at some large distance $r'$ from the white dwarf surface. At the same time, waves with higher frequency $\nu''>\nu'$ propagate almost freely through this outer region, although they are damped as they reach the altitude $r''<r'$ from the stellar surface with a higher value of magnetic field.

Let us find the values of physical parameters matching the observational data on PSR 1957+20 with our eclipse model. Substituting first $\nu_{_d}=0.318$~GHz and then $\nu_{_d}=1.4$~GHz in equation~(\ref{nu_d_eclipse}), and using corresponding values of eclipse region radii at these frequencies $r=R_{e}(\nu_{_d})$ we can find that this range of radio frequencies is damped at the distances $0.46 - 0.76\,R_{\odot }$ from the stellar surface, if 
\begin{equation}
\frac{B_{c}R_{c}^{3}}{\gamma_{_p}\left( 1-\cos \theta \right) }\approx 50~R_{\odot }^{3}.  \label{ecl-cond-1}
\end{equation}
First assume that the companion star is a hydrogen white dwarf with $R_{c}=0.145\,R_{\odot }$. Speculating further that the mean Lorentz factor of the particles of companion's magnetospheric plasma are moderate $\gamma_{_p}\sim 10 - 100$ and the average angle is rather large, $\theta=60^\circ - 90^\circ$, we obtain that the magnetic field strength at white dwarf's surface $B_{c}\sim 2\times 10^{5} - 3\times 10^{6}$~G. Alternatively, if the companion is built by pure helium (so that $R_{c}=0.046\,R_{\odot }$), equation~(\ref{ecl-cond-1}) yields $B_{c}\sim 5\times 10^{6} - 10^{8}$ G for the same set of parameters.

Applying the above procedure to PSR J$2051-0827$ system we obtain that the band of radio frequencies up to about 2 GHz is damped in the companion's magnetosphere, if the following condition should is satisfied,
\begin{equation}
\frac{B_{c}R_{c}^{3}}{\gamma_{_p}\left( 1-\cos \theta \right) }\approx 5\,R_{\odot }^{3}.  \label{ecl-cond-2}
\end{equation}
Here it was taken into account that the waves with $\nu_{_d}=436$ MHz are damped at the distance $R_{c}=0.31\,R_{\odot }$. Assuming, as in the previous case, that $\gamma_{_p}\sim 10 - 100$ and $\theta =60^\circ - 90^\circ$, we can estimate that $B_{c}\sim 2\times 10^{4} - 4\times 10^{5}$ G for the hydrogen white dwarf ($R_{c}=0.135\,R_{\odot }$), and $B_{c}\sim 5\times 10^{5} - 10^{7}$ G for the helium white dwarf ($R_{c}=0.043\,R_{\odot }$).

For all the intermediate cases of white dwarfs made of various mixtures of He and H by mass (that is, various values of $X$, see Section~\ref{Nature of companions}) the surface magnetic field falls in the range between the above values of $B_{c}$ in both cases. Detailed spectral observations of both companion stars would determine the spectral class (hence, the chemical composition) of each white dwarf, as well as constrain the values of surface magnetic fields at their surfaces. In the frame of our model (eqs.~[\ref{ecl-cond-1},\ref{ecl-cond-2}]) this would provide an indirect measurement of average angles $\theta $ and mean Lorentz factors of companion's magnetospheric plasma particles. The latter, in turn, may specify the energy of the slowest (and the most numerous) particles of the pulsar wind. All in all, we see that both companion stars possess quite strong surface magnetic fields, hence we deal with so called magnetic white dwarfs, probably with mega-Gauss magnetic fields. Let us notice that such objects are quite common in the Galaxy \citep{lang92}.

The necessary condition of the cyclotron resonance development is that the characteristic timescale of the instability, $\tau_{_d}\equiv 1/|\Gamma_{d}|$, should be much less than the time waves take to escape the resonance region $\tau_{_e}\simeq 2R_{e}/c$. The decrement $\Gamma_{d}$ is given by equation~(\ref{Decr_X}) for X-mode, and equation~(\ref{Decr_OA}) for O and A-modes. Assuming a relatively broad distribution function for companion's magnetospheric plasma (so that $\gamma_{_p}\sim \gamma_{_T}$) the former can
be rewritten as follows
\begin{equation}
\Gamma_{d}^{\mathrm{X}}=-0.8\,{\frac{n_{_p}(a)}{\gamma_{_p}}}\left( \frac{
\nu_{_d}}{\mathrm{[GHz]}}\right) ^{-1},  \label{ecl-decr}
\end{equation}
while the latter differs from equation~(\ref{ecl-decr}) by an angle-dependent multiplier $g\left( \theta \right)$ (eq.~[\ref{g(theta)}], see also Fig.~\ref{g(theta)-plot}). Thus, for pulsar to be invisible at some frequency $\nu_{_d}(R_{e}),$ the following two conditions should be satisfied at the corresponding distance $R_{e}$ from the companion's surface 
\begin{equation}
\frac{\tau_{_e}}{\tau_{_d}}=\frac{2\left| \Gamma_{d}\right| R_{e}}{c}\gg 1,
\label{ecl-damp-cond} 
\end{equation}
\begin{equation}
\frac{2\left| \Gamma_{d}\right| R_{e}}{c}g\left( \theta \right) \gg 1
\label{ecl-damp-cond-OA}
\end{equation}

Combining equations~(\ref{ecl-decr}) and (\ref{n_ecl}), expressing $R_{e}$ from equation~(\ref{nu_d_eclipse}) by use of first equation~(\ref{ecl-cond-1}) and then equation~(\ref{ecl-cond-2}), we can rewrite the condition for X-mode damping (eq.~[\ref{ecl-damp-cond}])  as 
\begin{equation}
6\times 10^{-2}\,\frac{\kappa }{\gamma_{_p}}\left( \frac{\nu_{_d}}{\mathrm{[GHz]}}\right) ^{-4/3}\gg 1  \label{ecl-damp-cond1}
\end{equation}
in the magnetosphere of the companion to PSR B$1957+20$, and as
\begin{equation}
3\times 10^{-2}\,\frac{\kappa }{\gamma_{_p}}\left( \frac{\nu_{_d}}{\mathrm{[GHz]}}\right) ^{-4/3}\gg 1  \label{ecl-damp-cond2}
\end{equation}
in that of companion to PSR J$2051-0827$, respectively. According to equation~(\ref{ecl-damp-cond-OA}), the corresponding conditions for O and A modes in each case differ from equations~(\ref{ecl-damp-cond1}) and (\ref{ecl-damp-cond2}) by the angle-dependent multiplier $g(\theta )$ (eq.~[\ref{g(theta)}]) on the left-hand side of these inequalities.

\vskip 12pt~\scalebox{0.5}{\includegraphics{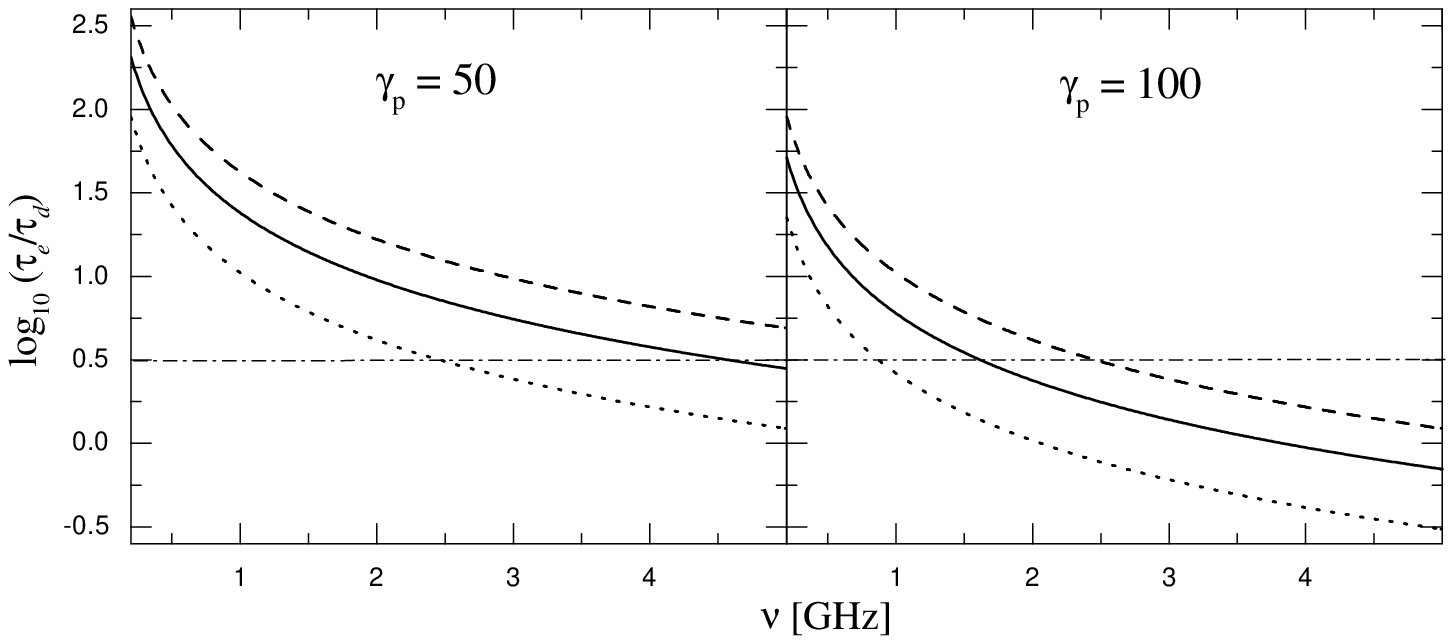}}~\figcaption[f4.eps]{Eclipse condition for PSR B$1957+20$ (eq.~[\ref{ecl-damp-cond1}]). See explanations in the text. \label{eclipse-condition-1}}~\vskip 2pt
\vskip 12pt~\scalebox{0.5}{\includegraphics{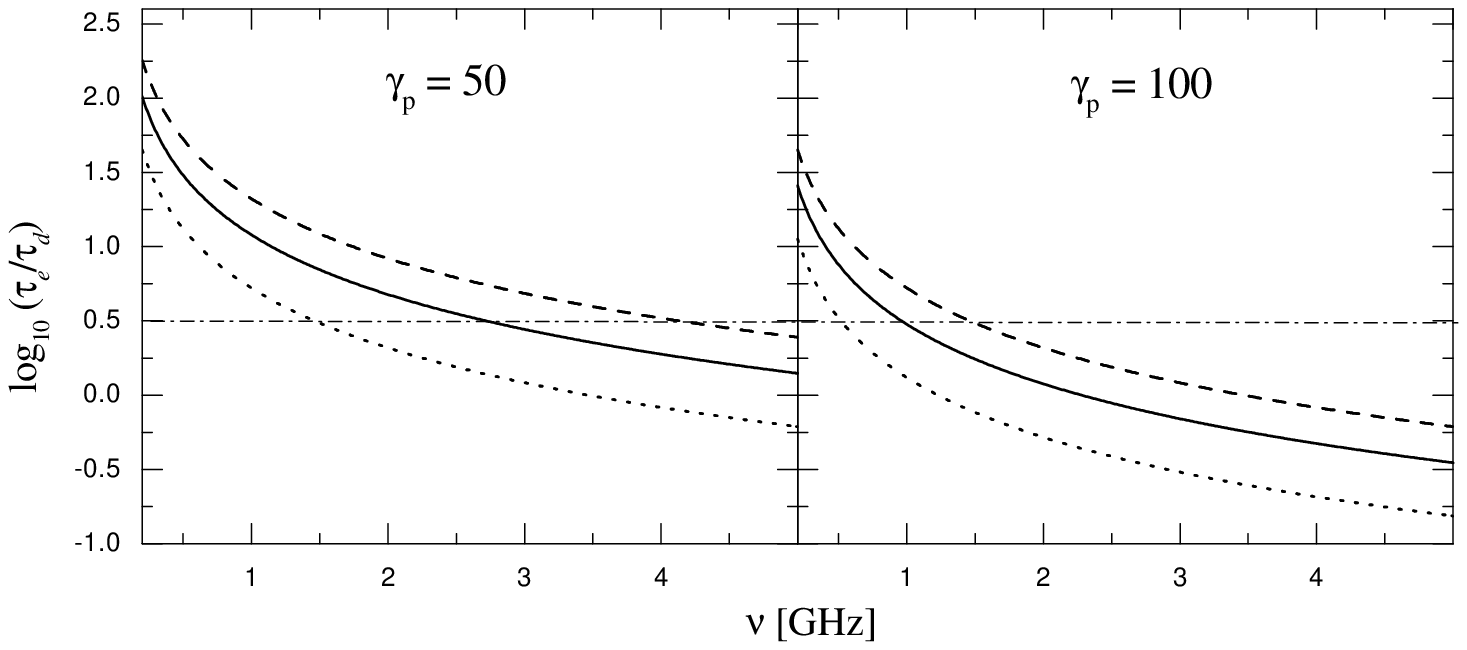}}~\figcaption[f5.eps]{Eclipse condition for PSR J$2051-0827$ (eq.~[\ref{ecl-damp-cond2}]). See explanations in the text. \label{eclipse-condition-2}}~\vskip 2pt

Figs.~\ref{eclipse-condition-1} and \ref{eclipse-condition-2} display damping conditions versus frequency in both cases. The solid lines correspond damping of X-mode at $\gamma_{_p}=50$ and $\gamma_{_p}=100$, and the dashed lines represent damping condition of coupled modes for the same values of mean Lorentz factor and the angle values $\theta =60^\circ$ and $\theta =80^\circ$. It is clear that all the intermediate cases fall between the curves presented in these figures. The dashed horizontal line corresponds to the characteristic value $\tau_{_e}/\tau_{_d}=3$ at which the intensity of waves drops by about $e^{2\tau_{_e}/\tau_{_d}}\approx 400$ times.

It is seen in Figure~\ref{eclipse-condition-1} that in the case of PSR B$1957+20$ the damping conditions are well satisfied for all the plasma wavemodes if $\gamma_{_p}=50$ (hence, $\kappa =2\times 10^{4}$). So damping is very strong and the pulsar radio emission certainly cannot reach an observer for this set of plasma parameters. However, the calculations corresponding $\gamma_{_p}=100$ (hence, $\kappa =10^{4}$) indicate that for such parameters the pulsar emission (at least one of the modes) may appear visible at relatively high frequencies during a low-frequency eclipse. However, the present lack of high-frequency eclipse observations of this pulsar do not allow us to judge strictly parameters of the magnetospheric plasma of its companion.

Figure~\ref{eclipse-condition-2} shows that the set of parameters $\gamma_{_p}=100$, $\kappa =10^{4}$ and $\theta =60^\circ - 80^\circ$ are certainly ruled out in the case of PSR J$2051-0827$, because the predictions are not consistent with the observational data from this binary system. However, it is seen that for $\gamma_{_p}=50$ (hence, $\kappa=2\times 10^{4}$) some critical frequency ($\nu_{_d}\approx 1.4$~GHz) may exist at which O and A-modes propagating with large angle $\theta \approx 80^\circ$ to the magnetic field escape from the eclipse region without significant loss, while X-mode is still strongly damped. It means that the pulsar remains visible at this frequency, although only the waves with one polarization can be detected. If such a situation appears at the edge of the eclipse region (where the angle $\theta$ varies in quite a small range), one can observe significant linear polarization at eclipse ingress and egress through quite a large frequency range (up to about 3~GHz, according to Fig.~\ref{eclipse-condition-2}). On the other hand, nonstationary processes taking place in the companion's magnetospheric plasma (such as, e.g., conal or cyclotron instabilities) may induce significant variations of plasma density in the companion's magnetospheric plasma. This may affect fulfillment of damping conditions, resulting in the eventual cyclotron damping of all the wavemodes at the critical frequency $\nu_{_d}\approx 1.4$~GHz. Thus, one observes an alternating eclipse at this frequency. Indeed, as it was mentioned in Section~\ref{PSR J$2051-0827$}, PSR J$2051-0827$ was detected at 1.4 GHz throughout the low-frequency eclipse region in about half of observing sessions \citep{stappers96}. Another reason of such an absence of eclipse at frequencies $\geq 1.4$~GHz could be a misalignment of the binary orbital plane with an observer's line of sight. Indeed, even at moderate inclination angle it is possible that observer misses completely the narrow
high-frequency ``eclipsing spot'' (with the radius $R_{e}(1.4~\mathrm{GHz})\approx 0.22\,R_{\odot }$), while at lower frequencies the pulsar still remains eclipsed at the same orbital phases (see Figs.~\ref{eclipse-geometry} and \ref{eclipse-spots}).

\section{Discussion}
We acknowledge that the model presented above is to some extent idealized. This regards, first of all, the energy distribution of plasma particles, which we assumed to be stable and constant with the altitude. In fact, a more realistic approach would imply that more energetic particles accumulate at larger distances from the stellar surface, whereas mean Lorentz factors of plasma particles decrease closer to the star, due to substantial synchrotron losses in the stronger magnetic field there. It is clear that this should affect calculation of both damping frequency (eq.~[\ref{nu_d_eclipse}]) and decrement (eq.~[\ref{ecl-decr}]). Indeed, according to our model, the eclipse length scales with the frequency as $\Delta t_{e}\propto  \nu^{-0.33}$. On the other hand, observations give $\Delta t_{e}\propto  \nu^{-0.4\pm 0.1}$ for PSR B$1957+20$ and $\Delta t_{e}\propto  \nu^{-0.15}$ for PSR J$2051-0827$, respectively. Such a deviation of the eclipse pattern from the theoretical prediction in both cases would result from the aforementioned variation of plasma distribution function with the altitudes from the stellar surfaces.

Another factor which can have a strong impact on the eclipse process described above is magnetic field orientation and its configuration. One can speculate about the orientation of the companion's magnetic field with respect to the star itself, and its magnetic axis could be inclined to the
rotation axis with an arbitrary yet constant angle (see Fig.~\ref{eclipse-geometry}). On the other hand, the long history of tidal interaction with the neutron star in such close binary systems as PSRs B$1957+20$ and J$2051-0827$ are, would align the angular momenta of the white dwarf and the neutron star with the binary orbital angular momentum. The same would also bring to the situation when the companion faces the pulsar always with one side (like the moon to the Earth), a fact apparently confirmed by the optical observations of both systems (see the discussion below). This implies that the companion's magnetosphere in both cases is strictly oriented with respect to the pulsar wind, as well as its radio emission cone (the latter being beamed around the pulsar magnetic axis ${\mathbf \mu }$). This, according to our model, would provide relative stability of the eclipse pattern, excluding eclipse variations associated with the eventual precession of the companion's magnetic axis around its rotation rotaion axis.

Regarding the configuration of companion's magnetic field, we assumed the latter being purely dipolar, although one can hypothesize that the global magnetic field of the white dwarf is modified by either a substantial contribution of local magnetic fields close to the stellar surface or, e.g., a quadrupole component. Naturally, such a change of the magnetic field value and direction should affect waves damping in the component's magnetosphere, especially at high radio frequencies (which are damped closer to the stellar surface).

It is worth estimating how the pulsar wind can influence the companion's magnetic field structure in the eclipse region. For that it is necessary to calculate the pulsar wind magnetic pressure 
\begin{equation}
w_{p}\equiv \frac{B_{p}^{2}}{8\pi }=\frac{\pi }{c}\frac{I}{R^{2}}\frac{\dot{P}}{P^{3}}  \label{w_p}
\end{equation}
at the distance of companion $a$ (here $I\approx 10^{45}$ g cm$^{2}$ is the neutron star moment of inertia and $R$ is a distance from the pulsar), and to compare it with the companion's magnetic pressure
\begin{equation}
w_{c}\equiv \frac{B_{c}^{2}}{8\pi }=4\times 10^{-2}B_{c}^{2}\left( \frac{
R_{c}}{r}\right) ^{6},  \label{w_c}
\end{equation}
in the eclipse region. For the parameters of each of the pulsars (see Section~\ref{Introduction}) equation~(\ref{w_p}) yields $w_{p}(a_{1})\approx 14~\mathrm{erg~cm^{-3}}$ and $w_{p}(a_{2})\approx 3~\mathrm{erg~cm^{-3}}$, respectively. On the other hand, at the distances from the white dwarf's surface corresponding to low-frequency eclipse region in each case ($R_{e}(318~\mathrm{MHz})=0.76\,R_{\odot }$ and $R_{e}(436~\mathrm{MHz})=0.31\,R_{\odot }$, respectively), we find that for the typical values given by equations~(\ref{ecl-cond-1}) and (\ref{ecl-cond-2}) the magnetic pressure $w_{c1}=1.3\times 10^{6}~\mathrm{erg~cm^{-3}}$ and $w_{c2}=2.8\times 10^{6}~\mathrm{erg~cm^{-3}}$, respectively. We see that $w_{p}\ll w_{c}$ in the eclipse region in both cases, so it can be said that the eclipse region of white dwarf's magnetosphere does not ``feel'' the pulsar wind at all, and the only (yet very important) role of the latter is to supply this magnetosphere with relativistic charged particles.

Moreover, one can estimate that the companion's magnetosphere dominates the pulsar wind magnetic field in the vast expanse of both binary systems. For PSR B$1957+20$ the ``inner front'' of the companion's magnetosphere, that is, the point between the white dwarf and the neutron star where $w_{p}\approx w_{c}$, is located at the distance $r_{\mathrm{inner}}\approx 2.3\,R_{\odot}$ from the companion, i.e. only about $0.2\,R_{\odot }$ from the pulsar surface (note that this distance is still equivalent to about $1800$ light cylinder radii of this pulsar). The latter means that the pulsar wind particles soon fall under control of the companion's magnetic field. At the same time, companion's magnetoitail is stretched up to $r_{\mathrm{outer}}\approx 8.3\,R_{\odot }$ away from the white dwarf. In the transverse direction companion's magnetosphere takes over pulsar wind up to $r_{\perp }\approx 7.5\,R_{\odot }$. The corresponding dimensions of PSR J$2051-0827$ companion's magnetosphere yield $r_{\mathrm{inner}}\approx 0.97\,R_{\odot },$ $r_{\mathrm{outer}}\approx 5.8\,R_{\odot }$ and $r_{\perp} \approx 5.4\,R_{\odot }$. We see that in this binary system pulsar wind particles start moving in the companion's field after they flee from the pulsar
magnetosphere on the distances more than about its 10 light cylinder radii.

Let us finish the discussion by a few remarks concerning possible physical mechanisms of the observed high-energy emission from these remarkable astrophysical systems. As it was mentioned in Section~\ref{Introduction}, highly variable optical counterparts of the companion stars were observed in both cases, and $X$-rays were detected from PSR B$1957+20$. It is out of the question that the pulsar is the only source of $X$-ray energy, but there are still lots of uncertainties in the explanation of the actual physical mechanism. There are no data so far on the $X$-ray spectrum, and it is still even not clear whether these $X$-rays are emitted by the millisecond pulsar itself, or they originate due to physical processes taking place somewhere between the pulsar and the white dwarf. \citet{thompson95} points to the fact that the large spin-down luminosity of PSR B$1957+20$, taking into account the radius of the companion, gives the energy flux at the companion's orbit which is seven times more than the one existing at the surface of our sun. In the model of \citet{artav93} the pulsar wind, shocked by its interaction with the wind off the companion, emits $X$-rays and $\gamma$-rays with a soft $X$-ray luminosity comparable to that observed. A large fraction of this radiation is then absorbed by the companion's surface facing the pulsar, due to the combination of relativistic beaming and the proximity of the companion to the shock. They show that the observed luminosity of $X$-rays can easily power also companion's optical emission.

To the best of our knowledge, there is so far no spectroscopic data on the optical counterparts of the companion stars in both binaries. Let us first assume that the optical emission will be found to be nonthermal. The fastest ``beam'' particles of the pulsar wind with $\gamma_{_b}\sim 10^{5 - 6} $ can reach the strong magnetic field at low altitudes from the stellar surface of the white dwarf, retaining their ultrarelativistic energies. Estimations show that such particles, gyrating around magnetic field lines, can emit a synchrotron emission falling into optical band. This might occur very close to the stellar surface, at distances from the companion's center that compare with the companion's optical sizes calculated from the observations (see Section~\ref{Introduction}).

Another mechanism can be suggested to explain the companion's high-energy radiation if the latter proves to be of thermal nature. Relativistic electrons and positrons trapped by the magnetic field of the white dwarf should ``bounce'' between higher-field points near the North and South poles (in the same manner as in the ordinary magnetic trap or in the Earth's magnetosphere) and slowly precess around the star due to $\nabla B$ and curvature drift. The most energetic particles with the ratio $v_{\parallel}/v_{\perp }$ at white dwarf's equatorial plane 
\begin{equation}
\frac{v_{\parallel }}{v_{\perp }}>\left( \frac{B_{\mathrm{max}}}{B_{\mathrm{min}}}-1\right) ^{1/2}  \label{trap}
\end{equation}
are in a so called ``loss cone'', and should be ``poured out'' on the magnetic poles. It can be demonstrated that broad polar regions of the companion star will be heated up to temperatures high enough to power the thermal $X$-ray emission. If this happens, the thermal optical emission can be just a long-wavelength tail of the Planck $X$-ray spectrum.

We would like to notice that the same physical mechanism was found to operate in the case of enigmatic pulsar Geminga \citep{gkm98}. Namely, it appeared that the cyclotron damping of radio waves in the magnetosphere of this pulsar, combined with the almost aligned geometry, leads to the apparent absence of its emission at radio frequencies higher than about 100~MHz. Moreover, we found that the cyclotron resonance has a strong impact on the general problem of the escape of radio waves from pulsar magnetospheres \citep{k99}. We plan to develop the eclipse model in forthcoming papers, where we expect to achieve better agreement with the observations and also include other eclipsing binary systems into the framework of our model.

\acknowledgements{We thank D. Lorimer for critical reading of the manuscript and many useful comments. The work is supported in part by the KBN Grants 2 P03D 015 12 and 2 P03D 003 15 of the Polish State Committee for Scientific Research. G. M. and D. K. acknowledge also support from the INTAS Grant 96-0154.}

\end{document}